\documentclass[12pt,reqno]{amsart}
\usepackage{amsfonts}
\usepackage{amssymb,amsmath}
\usepackage{amscd}
\usepackage{hyperref}
\usepackage{graphicx}
\usepackage{epsfig}

\newtheorem{example}{Example}[section]

\numberwithin{equation}{section}
\input{tcilatex}

\begin{document}
\title[A New Coding/Decoding Algorithm using Fibonacci Numbers]{A New
Coding/Decoding Algorithm using Fibonacci Numbers}
\author[N. TA\c{S}]{N\.{I}HAL TA\c{S}*}
\address{Bal\i kesir University\\
Department of Mathematics\\
10145 Bal\i kesir, TURKEY}
\email{nihaltas@balikesir.edu.tr}
\thanks{*Corresponding author: N. TA\c{S}\\
Bal\i kesir University, Department of Mathematics, 10145 Bal\i kesir, TURKEY%
\\
e-mail: nihaltas@balikesir.edu.tr}
\author[S. U\c{C}AR]{S\"{U}MEYRA U\c{C}AR}
\address{Bal\i kesir University\\
Department of Mathematics\\
10145 Bal\i kesir, TURKEY}
\email{sumeyraucar@balikesir.edu.tr }
\author[N. YILMAZ \"{O}ZG\"{U}R]{N\.{I}HAL YILMAZ \"{O}ZG\"{U}R}
\address{Bal\i kesir University\\
Department of Mathematics\\
10145 Bal\i kesir, TURKEY}
\email{nihal@balikesir.edu.tr}
\author[\"{O}. \"{O}ZTUN\c{C} KAYMAK]{\"{O}ZNUR \"{O}ZTUN\c{C} KAYMAK}
\address{Bal\i kesir University\\
10145 Bal\i kesir, TURKEY}
\email{oztunc@balikesir.edu.tr}
\date{}

\begin{abstract}
In this paper we present a new method of coding/decoding algorithms using
Fibonacci $Q$-matrices. This method is based on the blocked message
matrices. The main advantage of our model is the encryption of each message
matrix with different keys. Our approach will not only increase the security
of information but also has high correct ability.
\end{abstract}

\maketitle

\section{Introduction and Background}

\label{intro}

It is well known that the Fibonacci sequence is defined by 
\begin{equation}
F_{n}=F_{n-1}+F_{n-2}\text{ with }n\geq 2\text{,}  \label{eqn1}
\end{equation}%
with the initial terms $F_{0}=0,$ $F_{1}=1$. The Fibonacci $Q$-matrix is
defined in \cite{gould} and \cite{hoggat} as follows:%
\begin{equation*}
Q=\left( 
\begin{array}{cc}
1 & 1 \\ 
1 & 0%
\end{array}%
\right) .
\end{equation*}%
From \cite{stakhov 1999} and \cite{stakhov 2006}, we known that the $n.$th
power of the Fibonacci $Q$-matrix is of the following form:%
\begin{equation*}
Q^{n}=\left( 
\begin{array}{cc}
F_{n+1} & F_{n} \\ 
F_{n} & F_{n-1}%
\end{array}%
\right) .
\end{equation*}

In recent days information security becomes a more important matter in terms
of data transfer over communication channel. So coding/decoding algorithms
are of great importance to help in improving information security.
Especially, Fibonacci coding theory has been considered in many aspects (see 
\cite{basu}, \cite{prajat}, \cite{stakhov1999-2}, \cite{stakhov 2006}, \cite%
{Tarle} and \cite{Wang} for more details). For example, in \cite{stakhov
2006}, it was given a new coding theory using the generalization of the
Cassini formula for Fibonacci $p$-numbers and $Q_{p}$-matrices.

In this study we present a new coding/decoding algorithm using Fibonacci $Q$%
-matrices. The main idea of our method depend on dividing the message matrix
into the block matrices of size $2\times 2$. We use different numbered
alphabet for each message, so we get a more reliable coding method. The
alphabet is determined by the number of block matrices of the message
matrix. Our approach will not only increase the security of information but
also has high correct ability for data transfer over communication channel.

\section{The Blocking Algorithm}

\label{sec:1}

At first, we put our message in a matrix of even size adding zero between
two words and end of the message until we obtain the size of the message
matrix is even. Dividing the message square matrix $M$ of size $2m$ into the
matrices, named $B_{i}$ ($1\leq i\leq m^{2}$) of size $2\times 2$, from left
to right, we construct a new coding method.

Now we explain the symbols of our coding method. Assume that matrices $B_{i}$%
, $E_{i}$ and $Q^{n}$ are of the following forms:%
\begin{equation*}
B_{i}=\left( 
\begin{array}{cc}
b_{1}^{i} & b_{2}^{i} \\ 
b_{3}^{i} & b_{4}^{i}%
\end{array}%
\right) \text{, }E_{i}=\left( 
\begin{array}{cc}
e_{1}^{i} & e_{2}^{i} \\ 
e_{3}^{i} & e_{4}^{i}%
\end{array}%
\right) \text{ and }Q^{n}=\left( 
\begin{array}{cc}
q_{1} & q_{2} \\ 
q_{3} & q_{4}%
\end{array}%
\right) \text{.}
\end{equation*}%
The number of the block matrices $B_{i}$ is denoted by $b$. According to $b$%
, we choose the number $n$ as follows:%
\begin{equation*}
n=\left\{ 
\begin{array}{ccc}
3 & \text{,} & b\leq 3 \\ 
b & \text{,} & b>3%
\end{array}%
\right. \text{.}
\end{equation*}%
Using the choosen $n$, we write the following letter table according to $%
mod27$ (this table can be extended according to the used characters in the
message matrix). We begin the \textquotedblleft $n$\textquotedblright\ for
the first character.

\begin{equation*}
\begin{tabular}{|c|c|c|c|c|c|c|c|c|c|}
\hline
A & B & C & D & E & F & G & H & I & J \\ \hline
$n$ & $n+1$ & $n+2$ & $n+3$ & $n+4$ & $n+5$ & $n+6$ & $n+7$ & $n+8$ & $n+9$
\\ \hline
K & L & M & N & O & P & Q & R & S & T \\ \hline
$n+10$ & $n+11$ & $n+12$ & $n+13$ & $n+14$ & $n+15$ & $n+16$ & $n+17$ & $%
n+18 $ & $n+19$ \\ \hline
U & V & W & X & Y & Z & 0 &  &  &  \\ \hline
$n+20$ & $n+21$ & $n+22$ & $n+23$ & $n+24$ & $n+25$ & $n+26$ &  &  &  \\ 
\hline
\end{tabular}%
\end{equation*}

Now we explain the following new coding and decoding algorithms.

\textbf{Coding Algorithm (Fibonacci Blocking Algorithm)}

\textbf{Step 1.} Divide the matrix $M$ into blocks $B_{i}$ $\left( 1\leq
i\leq m^{2}\right) $.

\textbf{Step 2.} Choose $n$.

\textbf{Step 3. }Determine $b_{j}^{i}$ $\left( 1\leq j\leq 4\right) $.

\textbf{Step 4.} Compute $\det (B_{i})\rightarrow d_{i}$.

\textbf{Step 5.} Construct $F=\left[ d_{i},b_{k}^{i}\right] _{k\in
\{1,2,4\}} $.

\textbf{Step 6. }End of algorithm.

\textbf{Decoding Algorithm }

\textbf{Step 1.} Compute $Q^{n}$.

\textbf{Step 2.} Determine $q_{j}$ $(1\leq j\leq 4)$.

\textbf{Step 3. }Compute $q_{1}b_{1}^{i}+q_{3}b_{2}^{i}\rightarrow e_{1}^{i}$
$\left( 1\leq i\leq m^{2}\right) $.

\textbf{Step 4.} Compute $q_{2}b_{1}^{i}+q_{4}b_{2}^{i}\rightarrow e_{2}^{i}$%
.

\textbf{Step 5.} Solve $%
(-1)^{n}d_{i}=e_{1}^{i}(q_{2}x_{i}+q_{4}b_{4}^{i})-e_{2}^{i}(q_{1}x_{i}+q_{3}b_{4}^{i}) 
$.

\textbf{Step 6. }Substitute for $x_{i}=b_{3}^{i}$.

\textbf{Step 7.} Construct $B_{i}$.

\textbf{Step 8.} Construct $M$.

\textbf{Step 9.} End of algorithm.

In the following examples we give applications of the above algorithm for $%
b>3$ and $b\leq 3$, respectively.

\begin{example}
\label{exm1} Let us consider the message matrix for the message text
\textquotedblleft NIHAL HELLO\textquotedblright $:$%
\begin{equation*}
M=\left( 
\begin{array}{cccc}
N & I & H & A \\ 
L & 0 & H & E \\ 
L & L & O & 0 \\ 
0 & 0 & 0 & 0%
\end{array}%
\right) _{4\times 4}.
\end{equation*}%
\textbf{Coding Algorithm:}

\textbf{Step 1. }We can divide the message matrix $M$ of size $4\times 4$
into the matrices, named $B_{i}$ $\left( 1\leq i\leq 4\right) $, from left
to right, each of size is $2\times 2:$%
\begin{equation*}
B_{1}=\left( 
\begin{array}{cc}
N & I \\ 
L & 0%
\end{array}%
\right) \text{, }B_{2}=\left( 
\begin{array}{cc}
H & A \\ 
H & E%
\end{array}%
\right) \text{, }B_{3}=\left( 
\begin{array}{cc}
L & L \\ 
0 & 0%
\end{array}%
\right) \text{ and }B_{4}=\left( 
\begin{array}{cc}
O & 0 \\ 
0 & 0%
\end{array}%
\right) \text{.}
\end{equation*}

\textbf{Step 2.} Since $b=4\geq 3$, we choose $n=4$. For $n=4$, we use the
following \textquotedblleft letter table\textquotedblright\ for the message
matrix $M:$%
\begin{equation*}
\begin{tabular}{|l|l|l|l|l|l|l|l|l|l|l|}
\hline
$N$ & $I$ & $H$ & $A$ & $L$ & $0$ & $H$ & $E$ & $L$ & $L$ & $O$ \\ \hline
$17$ & $12$ & $11$ & $4$ & $15$ & $3$ & $11$ & $8$ & $15$ & $15$ & $18$ \\ 
\hline
\end{tabular}%
.
\end{equation*}

\textbf{Step 3.} We have the elements of the blocks $B_{i}$ $\left( 1\leq
i\leq 4\right) $ as follows:%
\begin{equation*}
\begin{tabular}{|l|l|l|l|}
\hline
$b_{1}^{1}=17$ & $b_{2}^{1}=12$ & $b_{3}^{1}=15$ & $b_{4}^{1}=3$ \\ \hline
$b_{1}^{2}=11$ & $b_{2}^{2}=4$ & $b_{3}^{2}=11$ & $b_{4}^{2}=8$ \\ \hline
$b_{1}^{3}=15$ & $b_{2}^{3}=15$ & $b_{3}^{3}=3$ & $b_{4}^{3}=3$ \\ \hline
$b_{1}^{4}=18$ & $b_{2}^{4}=3$ & $b_{3}^{4}=3$ & $b_{4}^{4}=3$ \\ \hline
\end{tabular}%
.
\end{equation*}

\textbf{Step 4.} Now we calculate the determinants $d_{i}$ of the blocks $%
B_{i}:$%
\begin{equation*}
\begin{tabular}{|l|}
\hline
$d_{1}=\det (B_{1})=-129$ \\ \hline
$d_{2}=\det (B_{2})=44$ \\ \hline
$d_{3}=\det (B_{3})=0$ \\ \hline
$d_{4}=\det (B_{4})=45$ \\ \hline
\end{tabular}%
.
\end{equation*}

\textbf{Step 5.} Using Step 3 and Step 4 we obtain the following matrix $F:$%
\begin{equation*}
F=\left( 
\begin{array}{cccc}
-129 & 17 & 12 & 3 \\ 
44 & 11 & 4 & 8 \\ 
0 & 15 & 15 & 3 \\ 
45 & 18 & 3 & 3%
\end{array}%
\right) .
\end{equation*}

\textbf{Step 6.} End of algorithm.

\textbf{Decoding algorithm:}

\textbf{Step 1.} It is known that%
\begin{equation*}
Q^{4}=\left( 
\begin{array}{cc}
5 & 3 \\ 
3 & 2%
\end{array}%
\right) \text{.}
\end{equation*}

\textbf{Step 2. }The elements of $Q^{4}$ are denoted by%
\begin{equation*}
q_{1}=5\text{, }q_{2}=3\text{, }q_{3}=3\text{ and }q_{4}=2\text{.}
\end{equation*}

\textbf{Step 3.} We compute the elements $e_{1}^{i}$ to construct the matrix 
$E_{i}:$%
\begin{equation*}
e_{1}^{1}=121\text{, }e_{1}^{2}=67\text{, }e_{1}^{3}=120\text{ and }%
e_{1}^{4}=99\text{.}
\end{equation*}

\textbf{Step 4. }We compute the elements $e_{2}^{i}$ to construct the matrix 
$E_{i}:$%
\begin{equation*}
e_{2}^{1}=75\text{, }e_{2}^{2}=41\text{, }e_{2}^{3}=75\text{ and }%
e_{2}^{4}=60\text{.}
\end{equation*}

\textbf{Step 5.} We calculate the elements $x_{i}:$%
\begin{eqnarray*}
(-1)^{4}(-129) &=&121(3x_{1}+6)-75(5x_{1}+9) \\
&\Rightarrow &x_{1}=15\text{.}
\end{eqnarray*}%
\begin{eqnarray*}
(-1)^{4}44 &=&67(3x_{2}+16)-41(5x_{2}+24) \\
&\Rightarrow &x_{2}=11\text{.}
\end{eqnarray*}%
\begin{eqnarray*}
(-1)^{4}0 &=&120(3x_{3}+6)-75(5x_{3}+9) \\
&\Rightarrow &x_{3}=3\text{.}
\end{eqnarray*}%
\begin{eqnarray*}
(-1)^{4}45 &=&99(3x_{4}+6)-60(5x_{4}+9) \\
&\Rightarrow &x_{4}=3\text{.}
\end{eqnarray*}

\textbf{Step 6.} We rename $x_{i}$ as follows$:$%
\begin{equation*}
x_{1}=b_{3}^{1}=15\text{, }x_{2}=b_{3}^{2}=11\text{, }x_{3}=b_{3}^{3}=3\text{
and }x_{4}=b_{3}^{4}=3\text{.}
\end{equation*}

\textbf{Step 7. }We construct the block matrices $B_{i}:$%
\begin{equation*}
B_{1}=\left( 
\begin{array}{cc}
17 & 12 \\ 
15 & 3%
\end{array}%
\right) \text{, }B_{2}=\left( 
\begin{array}{cc}
11 & 4 \\ 
11 & 8%
\end{array}%
\right) \text{, }B_{3}=\left( 
\begin{array}{cc}
15 & 15 \\ 
3 & 3%
\end{array}%
\right) \text{ and }B_{4}=\left( 
\begin{array}{cc}
18 & 3 \\ 
3 & 3%
\end{array}%
\right) \text{.}
\end{equation*}

\textbf{Step 8.} We obtain the message matrix $M:$%
\begin{equation*}
M=\left( 
\begin{array}{cccc}
17 & 12 & 11 & 4 \\ 
15 & 3 & 11 & 8 \\ 
15 & 15 & 18 & 3 \\ 
3 & 3 & 3 & 3%
\end{array}%
\right) =\left( 
\begin{array}{cccc}
N & I & H & A \\ 
L & 0 & H & E \\ 
L & L & O & 0 \\ 
0 & 0 & 0 & 0%
\end{array}%
\right) .
\end{equation*}

\textbf{Step 9.} End of algorithm.
\end{example}

\begin{example}
\label{exm2} Let us consider the message matrix for the message text
\textquotedblleft MATH\textquotedblright $:$%
\begin{equation*}
M=\left( 
\begin{array}{cc}
M & A \\ 
T & H%
\end{array}%
\right) _{2\times 2}.
\end{equation*}

\textbf{Coding Algorithm:}

\textbf{Step 1. }Since the size\textbf{\ }of message matrix M is\textbf{\ }$%
2\times 2$, we have only one block matrix $B_{1}=M.$

\textbf{Step 2.} Since $b=1<3$, we choose $n=3$. For $n=3$, we use the
following \textquotedblleft letter table\textquotedblright\ for the message
matrix $M:$%
\begin{equation*}
\begin{tabular}{|l|l|l|l|}
\hline
$M$ & $A$ & $T$ & $H$ \\ \hline
$15$ & $3$ & $22$ & $10$ \\ \hline
\end{tabular}%
.
\end{equation*}

\textbf{Step 3.} We have the elements of the blocks $B_{1}$ as follows$:$%
\begin{equation*}
\begin{tabular}{|l|l|l|l|}
\hline
$b_{1}^{1}=15$ & $b_{2}^{1}=3$ & $b_{3}^{1}=22$ & $b_{4}^{1}=10$ \\ \hline
\end{tabular}%
\text{.}
\end{equation*}

\textbf{Step 4.} Now we calculate the determinant $d_{1}$ of the block
matrix $B_{1}:$%
\begin{equation*}
\begin{tabular}{|l|}
\hline
$d_{1}=\det (B_{1})=84$ \\ \hline
\end{tabular}%
\text{.}
\end{equation*}

\textbf{Step 5.} Using Step 3 and Step 4 we obtain the following matrix $F:$%
\begin{equation*}
F=\left( 
\begin{array}{cccc}
84 & 15 & 3 & 10%
\end{array}%
\right) .
\end{equation*}

\textbf{Step 6.} End of algorithm.

\textbf{Decoding algorithm:}

\textbf{Step 1.} It is known that%
\begin{equation*}
Q^{3}=\left( 
\begin{array}{cc}
3 & 2 \\ 
2 & 1%
\end{array}%
\right) \text{.}
\end{equation*}

\textbf{Step 2. }The elements of $Q^{3}$ are denoted by%
\begin{equation*}
q_{1}=3\text{, }q_{2}=2\text{, }q_{3}=2\text{ and }q_{4}=1\text{.}
\end{equation*}

\textbf{Step 3.} We compute the element $e_{1}^{1}$ to construct the matrix $%
E_{1}:$%
\begin{equation*}
e_{1}^{1}=q_{1}b_{1}^{1}+q_{3}b_{2}^{1}=51\text{.}
\end{equation*}

\textbf{Step 4.} We compute the element $e_{2}^{1}$ to construct the matrix $%
E_{1}:$%
\begin{equation*}
e_{2}^{1}=q_{2}b_{1}^{1}+q_{4}b_{2}^{1}=33\text{.}
\end{equation*}

\textbf{Step 5.} We calculate the element $x_{1}:$%
\begin{eqnarray*}
(-1)^{3}(84) &=&51(2x_{1}+10)-33(3x_{1}+20) \\
&\Rightarrow &x_{1}=22\text{.}
\end{eqnarray*}

\textbf{Step 6.} We rename $x_{1}$ as follows$:$%
\begin{equation*}
x_{1}=b_{3}^{1}=22\text{.}
\end{equation*}

\textbf{Step 7. }We construct the block matrix $B_{i}:$%
\begin{equation*}
B_{1}=\left( 
\begin{array}{cc}
15 & 3 \\ 
22 & 10%
\end{array}%
\right) \text{.}
\end{equation*}

\textbf{Step 8.} We obtain the message matrix $M:$%
\begin{equation*}
M=\left( 
\begin{array}{cc}
15 & 3 \\ 
22 & 10%
\end{array}%
\right) =\left( 
\begin{array}{cc}
M & A \\ 
T & H%
\end{array}%
\right) .
\end{equation*}

\textbf{Step 9.} End of algorithm.
\end{example}

\section{A Computer Application}

\label{sec:2} To determine the verification of our coding method, in this
section we construct a computer algorithm. We creat the MATLAB codes for the
examples given in the previous section. So our blocking algorithm is checked
for $n=2$ and $n=4$ for different message texts, respectively. It can be
seen that the algorithm works errorless. Moreover, complex message texts are
solved correctly thanks to these algorithms. By a similar way, this
algorithm can be extended for convenient $n$. At first we define the
Fibonacci numbers in the algorithm then we give the following code segments
for $n=2$ (see Table \ref{Tab:1}).

\begin{table}[p]
\caption{The Algorithm for $n=2$}
\label{Tab:1}\centering
\par
\begin{tabular}{|c|}
\hline\hline
$a$=input('$a$=');\ $b$=input('$b$= ');\ $c$=input('$c$= ');\ $d$=input('$d$%
= '); \\ 
A= [$a$,$\ b$;$\ c$,$\ d$]; \\ 
F=[det(A), $a$; $c,d$]$;$\%End of Coding Algorithm \\ 
$fibf$(1) = 1;\ $fibf$(2) = 1; $p$=3; \  \\ 
while$\ $fibf($p-1)$ $<$ 1000 fibf($p$)=fibf(-1)+fibf($p$-2); \ $p$=$p$+1;\
end \\ 
Q = [fibf($n$+1),\ fibf($n$);\ fibf($n$),\ fibf($n$-1)]; \\ 
$e1$= fibf($n$+1)*a+fibf($n$)*b; $e2$= fibf($n$)*a+fibf($n$-1)*b; \\ 
$x$ = sym('$x$'); \\ 
eqn = (-1)\symbol{94}$n$*det(A) == ($e1$*(fibf($n$)*$x$+fibf($n$-1)*d))-($e2$%
*(fibf($n$+1)*$x$+fibf($n$)*d)); \\ 
solx = solve(eqn,'$x$'); \\ 
E =[a,b;solx,d] \%End of Decoding Algorithm \\ \hline
\end{tabular}%
\end{table}

The second algorithm seen in Table \ref{Tab:2} provides more faster solution
for complex text. The readers can verify these algorithms for varied values
of $n$.

\begin{table}[p]
\caption{The Algorithm for $n=4$}
\label{Tab:2}\centering%
\begin{tabular}{|c|}
\hline\hline
$a$=input('$a$=');\ $b$=input('$b$= ');\ $c$=input('$c$= ');\ $d$=input('$d$%
= '); \\ 
$e$=input('$e$ = ');\ $f$=input('$f$= ');\ $g$=input('$g$ = ');$h$=input('$h$
= '); \\ 
$i$=input('$i$= ');\ $j$=input('j = ');\ $k$=input('$k$ = ');$\ l$=input('$l$
= '); \\ 
$m$=input('$m$ = ');\ $n$=input('$n$ = ');\ $p$=input('$p$ = ');\ $r$=input('%
$r$ = '); \\ 
A = [$a,b,c,d;e,f,g,h;i,j,k,l;m,n,p,r$]; \\ 
M = [$a,b;e,f$]; N = [$c,d;g,h$]; O = [$i,j;m,n$]; P = [$k,l;p,r$]; \\ 
$d1$=det(M); $d2$=det(N); $d3$=det(O); $d4$=det(P); \\ 
F=[$d1,a,b,f;d2,c,d,h;d3,i,j,n;d4,k,l,r$]\%End of Coding Algorithm \\ 
$fibf$(1) = 1; $fibf$(2) = 1; $p$=3; \\ 
while\ $fibf$($p$-1) $<$ 1000 $fibf$($p$) = $fibf$($p$-1)+$fibf$($p$-2); ;
\\ 
$p$=$p$+1;end \\ 
de11=$fibf$($s$+1)*a+$fibf$($s$)*b; \\ 
e12= $fibf$($s$+1)*c+$fibf$($s$)*d; \\ 
e13= $fibf$($s$+1)*i+$fibf$($s$)*j; \\ 
e14= $fibf$($s$+1)*k+$fibf$($s$)*l; \\ 
e21= $fibf$($s$)*a+$fibf$($s$-1)*b; \\ 
e22= $fibf$($s$)*c+$fibf$($s$-1)*d; \\ 
e23= $fibf$($s$)*i+$fibf$($s$-1)*j; \\ 
e24= $fibf$($s$)*k+$fibf$($s$-1)*l; x = sym('$x$'); \\ 
eqn1 = (-1)\symbol{94}s*d1 == (e11*($fibf$($s$)*$x$+$fibf$($s$-1)*f))-(e21*($%
fibf$($s$+1)*$x$+$fibf$($s$)*f)); \\ 
eqn2 = (-1)\symbol{94}s*d2 == (e12*($fibf$($s$)*$x$+$fibf$($s$-1)*h))-(e22*($%
fibf$($s$+1)*$x$+$fibf$($s$)*h)); \\ 
eqn3 = (-1)\symbol{94}s*d3 == (e13*($fibf$($s$)*$x$+$fibf$($s$-1)*n))-(e23*($%
fibf$($s$+1)*$x$+$fibf$($s$)*n)); \\ 
eqn4 = (-1)\symbol{94}s*d4 == (e14*($fibf$($s$)*$x$+$fibf$($s$-1)*r))-(e24*$%
(fibf$($s$+1)*$x$+$fibf$($s$)*r)); \\ 
solx1 = solve(eqn1,'$x$'); solx2 = solve(eqn2,'$x$'); \\ 
solx3 = solve(eqn3,'$x$'); solx4 = solve(eqn1,'$x$'); \\ 
B1 = [$a,b$;solx1,$f];$ B2 = [$c,d$;solx2,$h$]; B3 = [$i,j;$solx3$,n$]; B4 =
[$k,l;$solx4,$r$]; \\ 
M=[$a,b,c,d$;solx1,$f$,solx2$,h;i,j,k,l;$solx3,$n$,solx4,$r$]\%End of
Decoding Algorithm \\ \hline
\end{tabular}%
\end{table}

\vskip0.5truecm

\section{Conclusion}

\label{sec:3} The main idea of our method depends on dividing the message
matrix into the block matrices of size $2\times 2$. We give an algorithm
using Fibonacci numbers and the numbers corresponding to each letter in used
alphabet changes for each new message matrix. We verify our new algorithm
with illustrative examples. Also our method is supported by a code in MATLAB
programme for $n=2$ and $n=4$. This algorithm can be improved for any $n$ in
the light of similar arguments.

\end{document}